\documentclass[11pt]{amsart}
\usepackage[centering]{geometry}                % See geometry.pdf to learn the layout options. There are lots.
\usepackage{graphicx}
\usepackage{amssymb}
\usepackage{epstopdf}
\usepackage{pdfpages}

\def\longrightharpoonup{\relbar\joinrel\rightharpoonup}
\def\longleftharpoondown{\leftharpoondown\joinrel\relbar}

\def\longrightleftharpoons{
  \mathop{
    \vcenter{
      \hbox{
	\ooalign{
	  \raise1pt\hbox{$\longrightharpoonup\joinrel$}\crcr
	  \lower1pt\hbox{$\longleftharpoondown\joinrel$}
	}
      }
    }
  }
}

\newcommand{\rates}[2]{\displaystyle
  \mathrel{\longrightleftharpoons^{#1\mathstrut}_{#2}}}

\newcommand{\diag}{\text{diag}}

\author{Micha{\l} Komorowski$^{1,*}$, Jacek  Mi\c{e}kisz$^{2}$,
Michael P.H. Stumpf$^{1,*}$
}

\title{Decomposing Noise in Biochemical Signalling Systems Highlights the Role of Protein Degradation}

\date{\today}                                           % Activate to display a given date or no date

\begin{document}
\pagestyle{plain}
\maketitle
\hspace{10mm}
{\small{
\begin{minipage}{\textwidth}
\noindent 1.Division of Molecular Biosciences, Imperial College London, UK\\
\noindent 2. Institute of Applied Mathematics and Mechanics,
University of Warsaw, Poland\\
\ \\{\small
\noindent $*$ Correspondence: {\it M.Komorowski@imperial.ac.uk, M.Stumpf@imperial.ac.uk }   }
\end{minipage}
}}
\begin{abstract}
The phenomena of stochasticity in biochemical processes have been intriguing life scientists for the past few decades. We now know that living cells take advantage of stochasticity in some cases and counteract stochastic effects in others.  The source of intrinsic stochasticity in biomolecular systems are random timings of individual reactions, which cumulatively drive the variability in outputs of such systems. Despite the acknowledged relevance of stochasticity in the functioning of living cells no rigorous method have been proposed to precisely identify sources of variability.  In this paper we propose a novel methodology that allows us to calculate contributions of individual reactions into the variability of a system's output. We demonstrate that some reactions have  dramatically different effects on noise than others.  Surprisingly, in the class of open conversion systems that serve as an approximate model of signal transduction, the degradation of an output contributes half of the total noise. We also demonstrate the importance of degradation in other relevant systems and propose a degradation feedback control mechanism that has the capability of an effective noise suppression. Application of our method to some well studied biochemical systems such as: gene expression, Michaelis-Menten enzyme kinetics, and the p53 system indicates that our methodology reveals an unprecedented insight into the origins of variability in biochemical systems. For many systems an  analytical decomposition is not available; therefore the method has been implemented as a Matlab package and is available from the authors upon request.
\end{abstract}

\ \\

Living cells need to constantly adapt to their changing environment. They achieve this through finely honed decision making and stress response machineries  that regulate and orchestrate the physiological adaptation to new conditions. In all studied genomes a large number of proteins have as their primary function the transfer and processing of such information. Such proteins are linked through a host of different mechanisms into biochemical circuits that perform a variety of information processing tasks including storage, amplification, integration of and marshalling the response to  environmental and physiological signals \cite{bray1995protein}. The functioning of these information processing networks depends on thermal or probabilistic encounters between molecules, resulting in a distortion of a transferred information that is best understood as noise.  Each reaction in the information processing machinery leads to an inevitable loss of information \cite{lestas467fundamental}.  Therefore, cell functions do not only rely on the necessity to make ``good" decisions, but also on appropriate ways to cope with the uncertainties arising from the ``noisy" signal transmission. To deal with the latter type of difficulty, we believe,  evolution equipped cells with reliable signal transduction systems by using less noisy reactions or reaction configurations, where needed \cite{silvanoise}. The question, however, which reactions, molecular species or parts of a network contribute most of the variability of a system or are responsible for most of the information loss has not gained much attention, except for some models of gene expression \cite{MichaelB-Elowitz08162002, Paszek, Paulsson2006, rausenberger2008quantifying}.  Origins of stochasticity in biochemical systems have been discussed and argued over at length, but a unified and robust mathematical framework to study this problem has been lacking. 
\par
Here we present a novel and general method to calculate contributions of individual reactions to the total variability of outputs of biochemical systems. This enables us to identify the  origins of cell-to-cell variability in dynamical biochemical systems. We derive a modified fluctuation-dissipation theorem which  enables us to determine how much of the total variance results from each of the system's reactions.  We then obtain some unexpected but general rules governing the noise characteristics of biochemical systems. In particular, we shall show  that in an arbitrary system with a sufficiently large number of molecules, degradation of the output (e.g. a reporter protein or a transcription factor) contributes  to the total variance of the system half the of the output's mean; for the important class of open conversion systems exactly half of the variance derives from the degradation step of the output signal. These results demonstrate that some reactions may be responsible for higher information loss than others; but our results also reveal that cells have the option of optimising biochemical network structures in order to avoid the most noisy reactions if necessary. 
Based on these results we propose a mechanism of controlled protein degradation based on a positive feedback that allows an arbitrary noise reduction resulting from the protein degradation.
\par
Below we first introduce the general framework for modelling  chemical reactions and derive a new method to decompose the noise in a biochemical system into contributions from different individual reactions.  Furthermore, two general properties governing noise are presented. Finally, we use biological examples of signal transduction systems to provide novel insights into the origins of variability. In particular, we decompose the variance of the  outputs  of linear transduction cascades and Michaelis-Menten enzyme kinetics. In addition,  for the oscillatory p53 system we show that stochasticity in p53 protein degradation is responsible for the variability in the oscillations' periodicity.

\section{Noise Decomposition}
We consider a general system of $n$ chemical species \cite{vanKapmen} inside a fixed volume  with $x=(x_1,\ldots ,x_n)^T$ denoting the number of molecules. The stoichiometric matrix
${S}=\{s_{ij}\}_{i=1,2\ldots n;\
j=1,2\ldots r}$ describes changes in the
population size due to $r$ different
reaction channels, where each $s_{ij}$
describes the change in the number of
molecules of type $i$ from $x_i$ to $x_i +
s_{ij}$ caused by an event of type $j$. The
probability that event $j$ occurs
during the time interval $[t,t+dt)$ equals
$f_j(x,t) dt$, where the $f_j(x,t)$ are called
 the transition rates. This
specification leads to a Poisson birth and death process described by the following stochastic equations \cite{ball2006asymptotic}

\begin{equation}\label{kurtz_x}
x(t) = x(0)+\sum_{j=1}^{r}S_{\cdot j}N_j(\int_0^t f_j(x,s)ds),
\end{equation}
where the $N_j$ are independent Poisson processes with rate 1.
\par
In order to define the contribution of the $j$th reaction $(j=1,..., r)$ to the variability of ${x}(t)$ we first define
$\langle {x}(t)\rangle_{(-j)}$ as the expectation of ${ x}(t)$ conditioned on the processes $N_1(t),...,N_{j-1}(t),N_{j+1}(t),...,N_r(t),$
 so that $\langle x(t)\rangle_{(-j)}$ is a random variable where timings of reaction $j$ have been averaged over all possible times, keeping all other reactions fixed. Therefore ${x}(t) - \langle {x}(t)\rangle_{|-j}$ is a random variable representing the difference between the native process ${ x}(t)$ and a time-averaged process of the $j$th reaction. Now the contribution of the $j$th  reaction to the total variability of ${x}(t)$ is 
\begin{equation}\label{contrib_def}
\Sigma^{(j)}(t)\equiv \langle ({ x}(t) - \langle { x}(t)\rangle_{|-j}) ({ x}(t) - \langle {x}(t)\rangle_{|-j})^T  \rangle,
\end{equation}
where $\langle\ldots \rangle$ denotes the temporal average over all $r$ reactions. This definition is similar to the one proposed in \cite{rausenberger2008quantifying} to quantify the contributions of promoter states and mRNA fluctuations, respectively, to the protein level variability.
\par
In general, it is difficult to calculate or study properties of equation (\ref{contrib_def}) using a Poisson birth and death process framework (\ref{kurtz_x}). Here instead we use the linear noise approximation (LNA), which allows us to model stochastic systems using Wiener processes driven $(W_j(s))$ instead of Poisson processes driven $(N_j(s))$  stochastic differential equations \cite{vanKapmen, JohanElf11012003, ja_LNA, Kurtz_Realation}. The LNA is valid if the number of interacting molecules is sufficiently large \cite{Kurtz_Realation} and decomposes  the system's state, $x(t)$, into a deterministic part $\varphi(t)$ and a stochastic part $\xi(t)$
\begin{equation}\label{dec_phi_xi}
{x}(t)=\varphi(t)+\xi(t);
\end{equation}
here $\varphi(t)$ and $\xi(t)$ are described by the deterministic and stochastic differential equations 
\begin{eqnarray}\label{ODE_phi}
\varphi(t)&=&\varphi(0)+\int_0^t S F(\varphi(s),s)ds\\\label{SDE_xi}
\xi(t)&=&\xi(0)+\int_0^t A(\varphi(s),s)\xi(s)ds \nonumber\\
&&+\sum_{j=1}^{r}S{\cdot j} W_j\left(  \int_0^t  \sqrt{{f}_j(\varphi(s),s)}    ds  \right),
\end{eqnarray}
respectively, and their coefficients are given by the following formulae
\begin{eqnarray}\label{F}
F(\varphi,t)=(f_1(\varphi,t), ..., f_r(\varphi,t) )^T\\  \label{A}
\left\{ A(\varphi,t) \right\}_{ik}=\sum_{j=1}^r s_{ij}
\frac{\partial f_j}{\partial \varphi_k}.
\end{eqnarray}
The LNA presents a simple way to compute contributions $\Sigma^{(j)}(t)$, and here we demonstrate how the total variance can be decomposed into the sum of individual contributions. We first write the explicit solution for the process $\xi$ as
\begin{equation}\label{sol_xi}
\xi(t)= \Phi(0,t)\xi(0) + \int_0^t \Phi(s,t) {E}(\varphi,s)dW(s),
\end{equation}
and ${E(\varphi,t)}=S\ \diag\left(\sqrt{f_1(\varphi,t))}, ...,\sqrt{(f_r(\varphi,t)}\right)$,
where
 $\Phi(s,t)$ is
the fundamental matrix of the non-autonomous
system of ordinary differential equations\footnote{In order to simplify notation, form now on we will write $A(t)$ and $E(t)$ instead of $A(\varphi, t)$ and $E(\varphi,t)$, respectively. }
\begin{equation}\label{fundamental}
\frac{d\Phi(s,t)}{dt}={A}(t)\Phi(s,t),\ \ \ \Phi(s,s)=I.
\end{equation}
Now it is straightforward to verify that
\begin{equation}\label{xi_j}
<\xi(t)>_{|-j}= \Phi(0,t)\xi(0) + \int_0^t \Phi(s,t)  {{E}}^{(-j)}(s)dW(s),
\end{equation}
where 
${{E}}^{(-j)}(t)=S^{(-j)}{{E}}(t)$, ${{E}}^{(j)}(t)=S^{(j)}{{E}}(t)$,  $S^{(-j)}=\{ s^{(-j)}_{lk}\}_{l=1,.., n, k=1,...,r}$,   $S^{(j)}=\{ s^{(j)}_{lk}\}_{l=1,.., n, k=1,...,r}$  and
$$
s^{(-j)}_{lk}=\left\{
\begin{array}{cc}
0& \text{ for } k=j\\ 
s_{lk}& \text{ for } k\neq j
\end{array} \right.\ \ \ \ \ \ \ \ \
s^{(j)}_{lk}=\left\{
\begin{array}{cc}
s_{lk}& \text{ for } k=j\\ 
0& \text{ for } k\neq j
\end{array} \right..
$$
From (\ref{sol_xi}) and (\ref{xi_j}) we have 
\begin{equation}
\xi(t) -  <\xi(t)>_{|-j}= \int_0^t \Phi(s,t) {{E}}^{(j)}(s)dW(s).
\end{equation}
With ${ x}(t)- \langle {x}(t)\rangle_{|-j} = \xi(t) -  \langle \xi(t)\rangle_{|-j}$ and the time derivative of  $\langle (\xi(t) -  \langle \xi(t)\rangle_{|-j}) (\xi(t) -  \langle \xi(t)\rangle_{|-j})^T \rangle $
we obtain for $\Sigma^{(j)}(t)$, 
\begin{equation}\label{Sigma_j}
\frac{d\Sigma^{(j)}}{dt}={{A}}(t)\Sigma^{(j)}+ \Sigma^{(j)} {{A}}(t)^T + {{D}}^{(j)}(t),
\end{equation}
with
\begin{equation}\label{D_j}
{{D}}^{(j)}(t)= {{E}}^{(j)}(t)({{E}}^{(j)}(t))^T.  
\end{equation}
This is, of course, analogous to the fluctuation dissipation theorem, with the exception that the diffusion matrix contains zeros at all entries not corresponding to the $j$th reaction.
Now the fact that  the total variance 
$$
\Sigma(t)=\langle (x(t) - \langle x(t)\rangle) (x(t) - \langle x(t)\rangle)^T  \rangle
$$ 
can be represented as the sum of individual contributions
\begin{equation}\label{Sigmaj_sum}
\Sigma(t)=\Sigma^{(1)}(t)+\ . . .\  +\Sigma^{(r)}(t),
\end{equation}
results directly from the decomposition of the diffusion matrix  ${{D}}(t)=\sum_{j=1}^{r} {{D}}^{(j)}(t)$ and
the linearity of the equation for $\Sigma(t)$, given by the standard fluctuation dissipation theorem 
\begin{equation}
\frac{d\Sigma}{dt}={{A}}(t)\Sigma+ \Sigma{{A}}(t)^T + {{D}}(t).
\end{equation}

\section{Mathematical Results}
With the decomposition (\ref{Sigmaj_sum}) it is in principle possible to detect reactions that make large contributions to the output variability of biochemical reactions. But even  simple systems, for  which analytic formulae exist, usually have complicated noise structures; we can nevertheless prove two general propositions that assign surprisingly substantial contributions to the degradation of an output signal. We formulate them in this section (proofs are in the {\em Appendix}) and illustrate them further below.
\par
\subsection{ Proposition 1}
Consider a general system such as described at the beginning of the {\em Noise Decomposition} section. In addition, assume that the deterministic component of the system $\varphi$ has a unique and stable stationary state (all eigenvalues of matrix $A$ have negative real parts). If $x_n$  is an output of this system being produced at rate $\sum_{i \neq n} k^{+}_i x_i$ and degraded in the $r$th reaction at rate $ k^{-}_n x_n$ then the contribution of  the output's degradation is equal to the half of its mean; more specifically, 
\begin{equation}
[\Sigma^{(r)}]_{nn}=\frac{1}{2} \langle x_n \rangle.
\end{equation}
\subsection{Proposition 2 (open conversion system)}  Now consider again a general system but assume that reaction rates are of mass action form and that only three types of reactions are allowed: 
\begin{description}
\item[production from source:]\hskip5mm$\emptyset  \xrightarrow{k^{+}_i } x_i$
\item[degradation:]\hskip20mm  $x_i  \xrightarrow{\ k^{-}_i }  \emptyset $
\item[conversion:]\hskip22mm $x_i  \xrightarrow{k_{ij} x_i}  x_j $. 
\end{description}
To satisfy the openness assumption each species can be created and degraded either directly or indirectly (via a series of preceding conversion reactions). As in {\it Proposition 1} let $x_n$  be the output of the system. Under this assumption the degradation of the output contributes exactly half of the total variance of the system's output, 
\begin{equation}
\left[\Sigma^{(r)}\right]_{nn}=\frac{1}{2}  \left[\Sigma\right]_{nn},
\end{equation}
where $r$ is again the index of the output's degradation reaction.
\par
\subsection{Remarks on propositions}
{\it Proposition 2} can be understood as a balance between production and degradation reactions. If we consider all reactions except $r$th as  producing species $x_n$, then production and degradation  contribute the same amount of noise. Usually, however, there is more than one production reaction, and therefore it is more convenient to interpret this result as the contribution of a single reaction.
\par
Both propositions indicate that a substantial part of noise is a result of the signal's degradation reaction. This observation is particularly interesting as open conversion systems are often good approximations to important biochemical reactions such as Michaelis-Menten enzymatic conversions or linear signal transduction pathways, which we discuss below; it suggests also that  controlled degradation is an effective mechanism to decrease the overall contribution to variation in protein levels.

\section{Biological Examples}
In this section we demonstrate how our methods can be used to study origins of noise in biochemical systems and present consequences of the above propositions.
\par
\subsection{Simple birth and death process} For illustrative purposes we start with a basic birth and death process, where molecules $x$ arrive at rate $k^{+}$ and degrade at rate $k^{-} x$. In the LNA this process can be expressed by the following stochastic differential equation
\begin{equation} 
dx=(k^{+}-k^{-} x(t)) dt+ \underbrace{\sqrt{k^{+}}dW_1}_{\text{birth noise}} +\underbrace{\sqrt{k^{-} \langle x(t) \rangle}dW_2}_{\text{death noise}}.
\end{equation}
The stationary distribution of this system is Poisson with the mean $\langle x \rangle=\frac{k^{+}}{k^{-}}$. Therefore
in the stationary state  death events occur at rate $ k^{-} \langle x \rangle$ which must be equal the birth rate $k^{+}$. The noise terms in the above equation are equal at stationarity indicating that contributions of birth and death reactions are equal. Using formula (\ref{Sigma_j}) it is straightforward to verify that the decomposition 
\begin{equation}
\Sigma= \underbrace{\frac{1}{2}\langle x \rangle}_{\text{birth noise}} +  \underbrace{\frac{1}{2} \langle x \rangle}_{\text{death noise}}
\end{equation}
holds indeed.
% The other possible interpretation of contributions of individual reaction is that it describes the level of noise if all the other reaction were deterministic. \\

\subsection{ Linear signal transduction cascades}  Here we first focus on conversion pathways, where molecules $x_1$ are  born at a certain rate $k^{+}_1$ and information is transmitted to molecules $x_n$ using (possibly reversible) conversion reactions. All molecules can degrade with arbitrary first order rates, and we assume that the final product $x_n$ is not the substrate of any conversion reaction. Under these assumptions, regardless of parameter values, length of the pathway, and degradation of intermediates, the degradation of the output $x_n$, rather unexpectedly, contributes exactly half of the variance of the signalling output, $x_n$.
On the other hand, if information is transmitted not using conversion reactions, but each species catalyses the creation of the subsequent species, i.e.  $x_i  \xrightarrow{k_{i\ i+1} x_i} x_i+ x_{i+1}$, then according to the {\it Proposition 1}, the contribution of the degradation of $x_n$ to the output's variance equals half of its mean regardless of the cascade length and parameter values.
\par
\begin{figure}
 \includegraphics[scale=1]{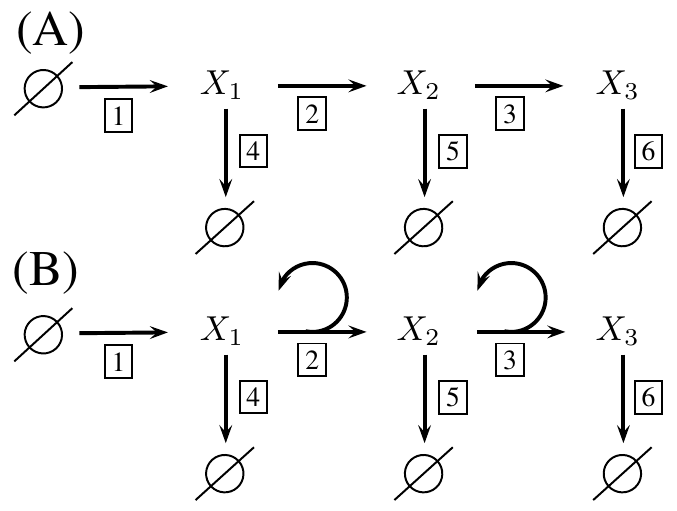}
\caption{Three step conversion (A) and catalytic (B) pathways. 
Rates $f_j$ for reactions $j=1,...,6$ have the form:  $f_1=k^{+}_1,  f_2=k^{+}_2 x_1, f_3=k_3^{+} x_2, f_4=k_4^{-}x_1, f_5=k^{-}_5x_5, f_6=k_6^{-} x_6$.
\label{Cascades_arrows} 
}\end{figure}

\begin{figure}
 \includegraphics[scale=0.3]{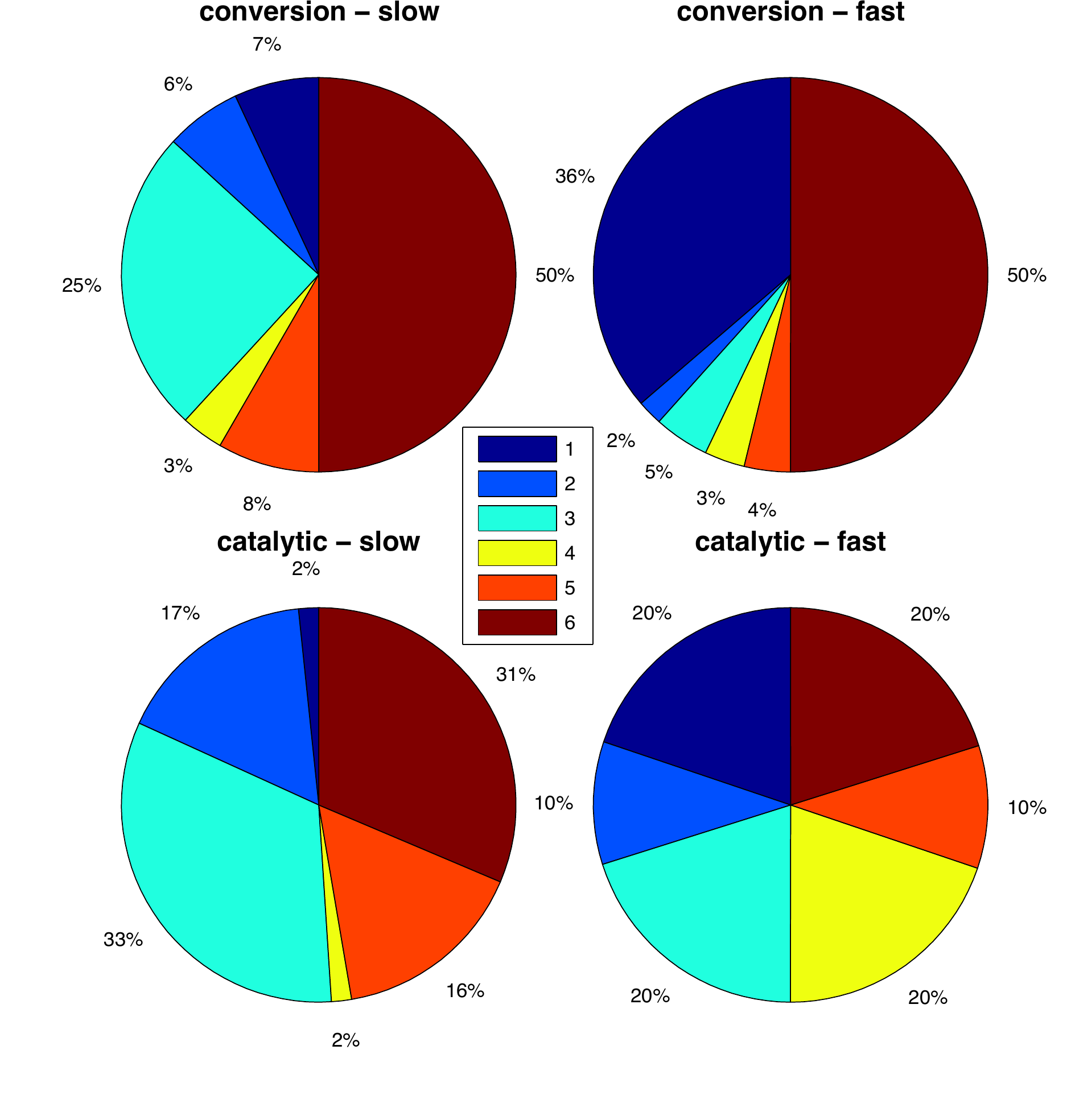}
 \caption{Variance contributions of the different reactions in the three steps linear conversion (top)  and catalytic cascades (bottom). Parameters for each case are given below.
Conversion slow: 
$k^{+}_1= 50, k^{+}_2= 1, k^{+}_3= 1, k^{-}_4=1, k^{-}_5=1, k^{-}_6=1$;
conversion fast: 
$k^{+}_1= 50, k^{+}_2= 10, k^{+}_3= 10, k^{-}_4=1, k^{-}_5=1, k^{-}_6=1$;
catalytic slow:
$k^{+}_1= 50, k^{+}_2= 0.1, k^{+}_3= 0.1, k^{-}_4=1, k^{-}_5=0.1, k^{-}_6=0.1$;
catalytic fast:
$k^{+}_1= 50, k^{+}_2= 10, k^{+}_3= 10, k^{-}_4=1, k^{-}_5=10, k^{-}_6=10$.
All rates are per hour.  \label{Cascades}    }
  \end{figure}

	
In order to study contributions of intermediate reactions, we considered cascades of the length three (see Figure \ref{Cascades_arrows}) and decomposed variances numerically for both conversion and catalytic cases. Decompositions are presented in Figure \ref{Cascades}. For conversion cascades we used two parameter sets corresponding to fast  and slow conversions to demonstrate that increasing the rates of intermediate steps decreases their contribution as might intuitively be predicted (top panel of Figure \ref{Cascades}). 
For catalytic cascades increasing the rates of reactions at step two and three (going from slow to fast) increases the contribution of reactions at the bottom of the cascades. For slow dynamics during steps two and three, the  relatively fast fluctuations at the start are filtered out (low pass filtering). If the dynamics of all species occur at similar times scales then these fluctuation can efficiently propagate downstream (bottom panel of Figure \ref{Cascades}). 
 \subsection{ Michaelis-Menten kinetics} 
\begin{figure}
\includegraphics[scale=0.3]{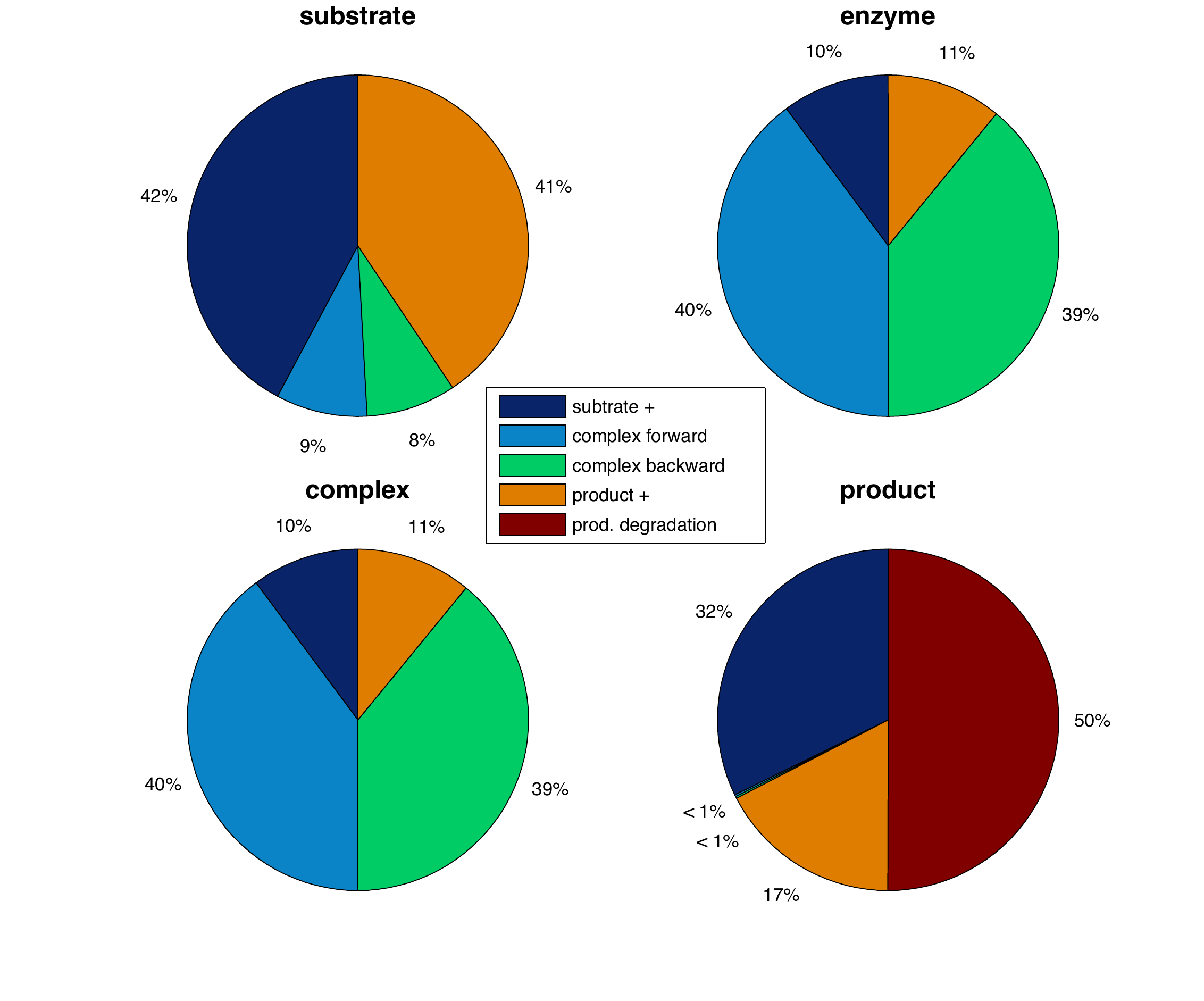}
\caption{ \label{michaelis}  
Variance decomposition for the four species of the Michaelis-Menten kinetics model. The following parameters were used: 
$k_0=0.1,\ k_1=50,\  k_2=1,\  k_b=20,\  k_d =0.1$. All rates are per hour.
 }
  \end{figure}

Michaelis-Menten enzyme kinetics are fundamental examples of biochemical reactions;
 substrate molecules ($S$) bind reversibly to enzyme molecules ($E$) with the forward rate constant $k_0$ and the backward rate constant $k_1$ to form a complex ($C$), which then falls apart into the enzyme and a product ($P$)  at rate $k_2$:  $S+E 
\rates{k_{ 0}S\cdot E}{k_{1}} C \xrightarrow{k_2} E+P$. To ensure existence of the steady state, we assume that substrate molecules arrive at rate $k_b$ and are degraded at rate $k_d$. At the unique steady state, if the concentration of an enzyme is large compared to the substrate, the system is well approximated by a set of mono-molecular reactions: 
$S \rates{k_{ 0}S\cdot E}{k_{1}} C \xrightarrow{k_2} P.$ Our theory therefore predicts that half of the noise in Michaelis-Menten kinetics operating in its linear range (abundant enzyme) is generated by degradation of the product molecules. 
In Figure \ref{michaelis} we have calculated the contributions of each of the four reactions to the variance of all  four species for the full model (without themonomolecular approximation). The contribution of the output degradation is as predicted by the monomolecular approximation.

\subsection{Gene expression} 
\begin{figure}
\includegraphics[scale=0.35]{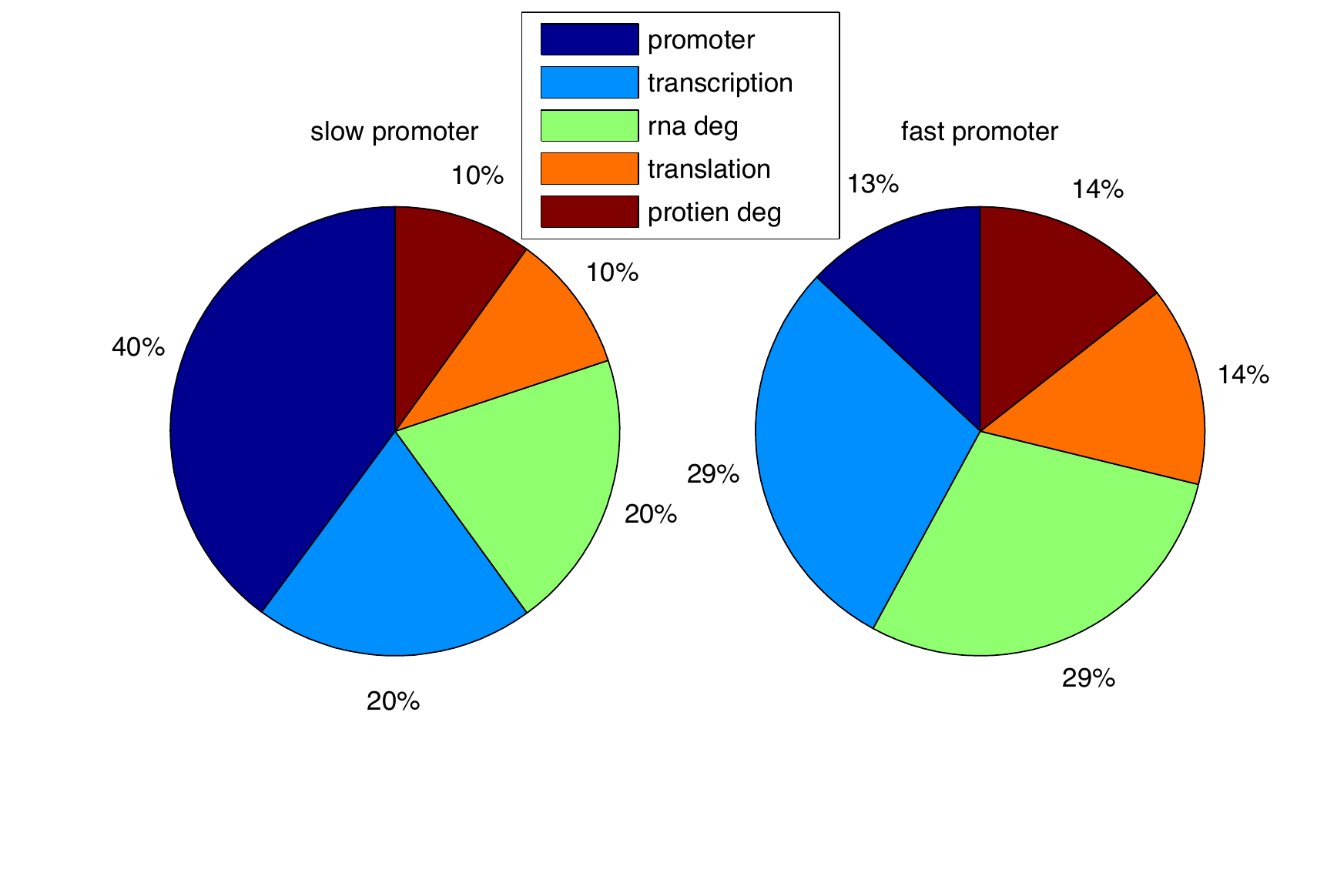}
 \caption{Variance contribution of the reactions involved in gene expression with fluctuating promoter states with transcription rate modelled as $k_1^{+}=\frac{Vy/H}{1+y/H}$, and $y$
being the stationary solution of $dy=(b-\gamma y)dt+\sqrt{2\gamma}dW$. The following parameters were used:  $V=100$, $H=50$, $k_2^{+}=2$ $k^{-}_1=0.44$, $k^{-}_2=0.55$.
For fast fluctuations we used $b=50$,$\gamma=1$, and for slow $b=0.5$, $\gamma=0.01$. All rates are per hour.
  \label{promoter}  }
  \end{figure}

The canonical example of a linear catalytic pathway is the gene expression process, which  can simply be viewed as the production of RNA ($x_1$) from source at the rate $k_1^{+}$, and production of protein ($x_2$) in a catalytic reaction at rate $k_2x_1$ together with first order degradation of both species at rates $k_1^{-}x_1$, $k_2^{-}x_2$.{\it Proposition 1} states that the part of the variance resulting from the protein degradation equals half of the protein mean. Moreover, formula (\ref{Sigma_j})  allows us to derive the complete decomposition 
\begin{equation}
V(x_2)=\underbrace{\frac{1}{2}\langle x_2\rangle }_{\text{prot. degradation}}+\underbrace{\frac{1}{2}\langle x_2\rangle}_{\text{translation}} + \underbrace{\frac{1}{2}\frac{k_p \langle x_2\rangle}{\gamma_r+\gamma_p}}_{\text{mRNA degradation}}+ \underbrace{\frac{1}{2}\frac{k_p\langle x_2\rangle}{\gamma_r+\gamma_p}}_{\text{ transcription}}.
\end{equation}
A similar decomposition was first presented by \cite{Paulsson2006, Paszek}, however, only into contributions resulting from fluctuating species and without theoretical explanation;  the latter was provided later by  \cite{rausenberger2008quantifying}. Below we investigate two extensions of the above model. First, we assume that the promoter can fluctuate between "on" and "off" states (similarly to \cite{Paszek, rausenberger2008quantifying}) and calculate contributions for different time-scales of these fluctuations. Second, we assume that the protein is a fluorophore that undergoes a two-step maturation process before it becomes visible (folding and joint cyclization with oxidation). Figure \ref{promoter} presents contributions for fast and slow promoter kinetics, showing that fast fluctuations are effectively filtered out (contributing 10 \%) but remain a substantial contributor when they are slow (contributing 40 \%). 
\par
\begin{figure}
 \includegraphics[scale=0.3]{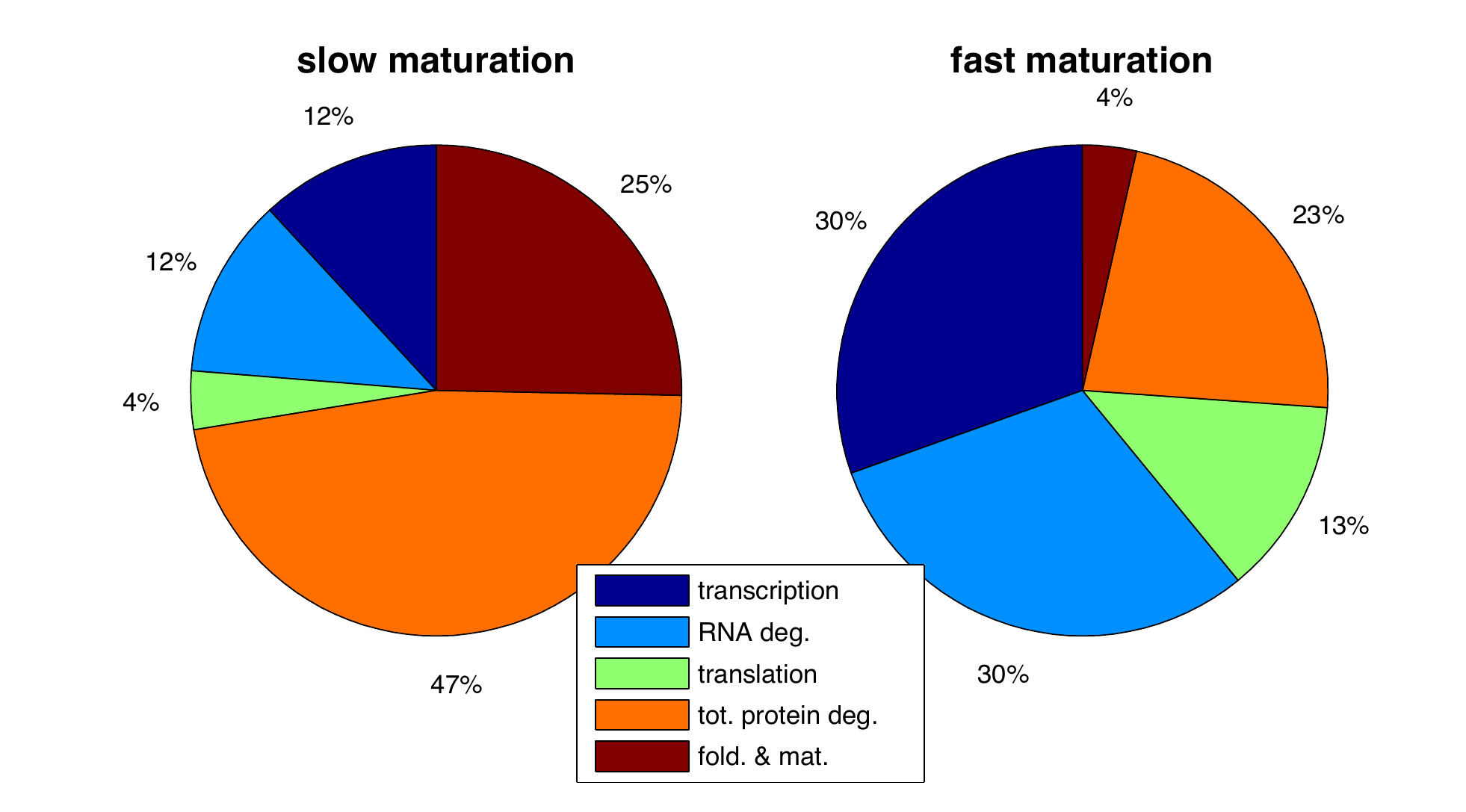}
  \caption{Variance contributions of the reactions involved in fluorescent protein maturation model for slow (left) and fast (right) maturing proteins. The following parameters were used:
 $k^{+}_1=50,\ k^{-}_1=0.44,\ k^{+}_2= 2,\ k^{-}_2=0.55$.  For slow maturation (average maturation time approx. 5 h.) we assumed folding and maturation rates to be $k_f=0.2$ and $k_m =1.3$, respectively. For fast maturation (average maturation time approx.  0.5 h) we set $k_f=2.48$, $k_m =13.6$. All rates are per hour.
  \label{Maturation}    }
  \end{figure}

Variability in gene expression is often measured by means of fluorescent proteins that undergo maturation before becoming visible for detection techniques; but the process of maturation \cite{tsien1998gfp} itself is subject to stochastic effects, and can thus contribute significantly to the observed variability.  We used typical parameters  \cite{bevis2002rapidly, tsien1998gfp} for fast and slow maturing fluorescent proteins and found that maturation contributed 4 and 25\%, respectively (Figure \ref{Maturation}) to the overall variability; here our method allows for the rigorous quantification compared to the previous qualitative observations of \cite{dong2008epm,komorowski2010using}. 

\subsubsection{Out-of-steady state systems: p53 model}
\begin{figure}
\includegraphics[scale=0.25]{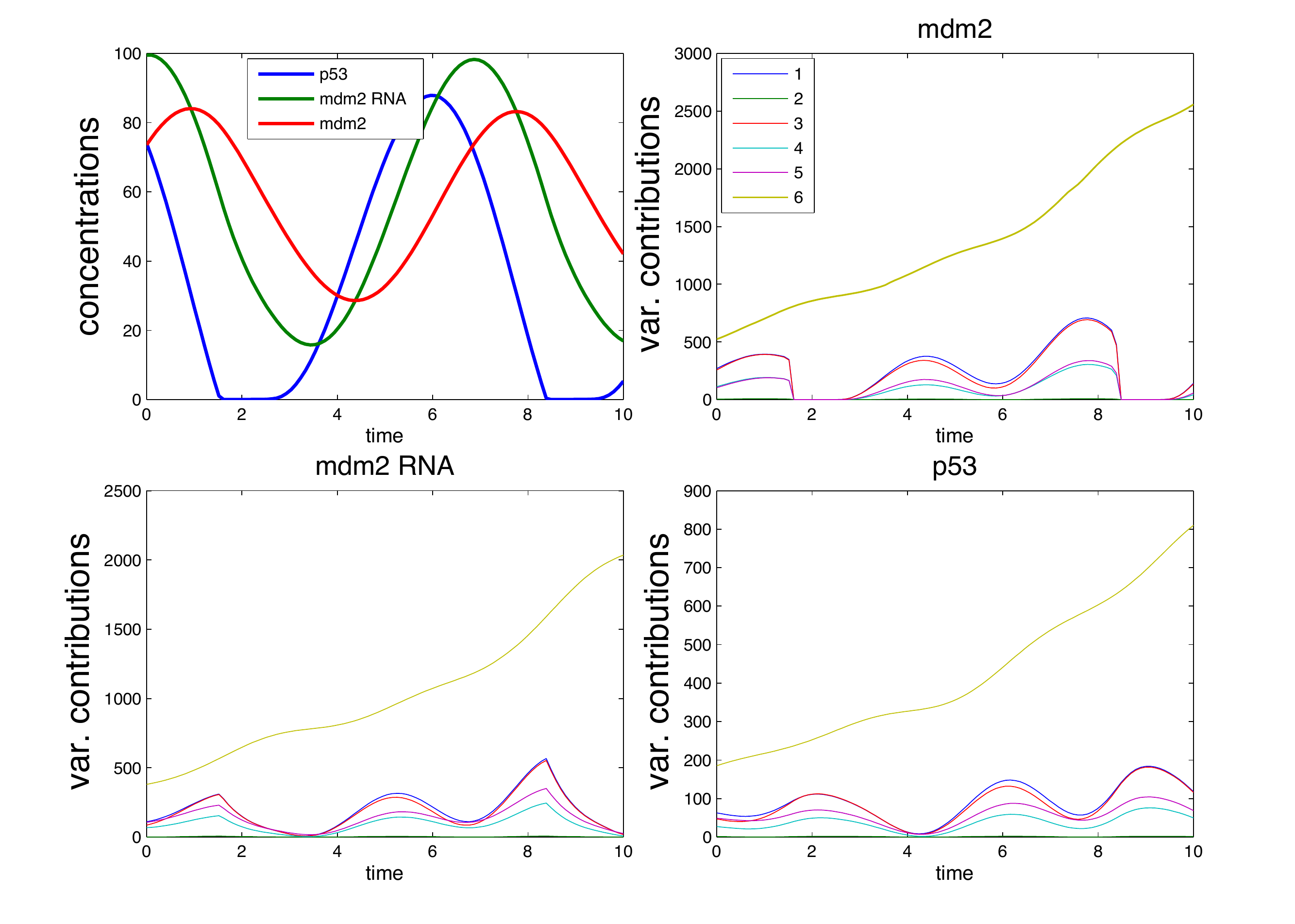}
 \caption{ \label{P53_contrib}  
 Trajectories (top left) and decomposition of variance into the contributions corresponding to each of the $6$ reactions for mdm2 (top right), mdm2 RNA (bottom left) and p53 (bottom right). We used the index $i$ of the parameters $k_i$ in equations (\ref{p53_1}-\ref{p53_3}) to denote reaction they describe. We used model and parameters  published in \cite{geva2006oscillations}.  Figure demonstrates accumulation of the noise contributed by degradation of p53 ($6$th reaction). 
The following parameters were used: $k_1= 90,\  k_2=0.002,\  k_3=1.7,\  H=0.01,\  k_4=1.1,\  k_5=0.8,\ k_6=0.8.$ All rates are per hour.}
  \end{figure}

In order to demonstrate that our methodology and predictions are valid also for more general out-of-steady state models we next focus on the p53 regulatory system which incorporates a feedback loop between the tumour suppressor p53 and the oncogene Mdm2, and is involved in the regulation of  cell cycle and the response to DNA damage. We consider the model introduced in \cite{geva2006oscillations} and later analysed in \cite{we_PNAS} that reduces the system to three  molecular species, p53, mdm2 precursor and  mdm2, denoted here by $\varphi_p, \varphi_{y_0}$  and $\varphi_y$. The model incorporates six reactions ($i=1,\ldots,6$), and its deterministic version can be written in the form of the following ordinary differential equations
\begin{eqnarray}\label{p53_1}
\dot{\varphi}_p&=& k_1-k_2 \varphi_p -k_{3}\varphi_y\frac{\varphi_p}{\varphi_p+H}\\
\dot{\varphi}_{y_0}&=&k_4 \varphi_p- k_5 \varphi_{y_0} \\\label{p53_3}
\dot{\varphi}_y&=& k_5 \varphi_{y_0} - k_6 \varphi_{y}.
\end{eqnarray}
Using equation (\ref{Sigma_j}) we calculated the contributions of all of the six reactions present in the model. Results are presented in Figure  \ref{P53_contrib}. The contributions of each reaction oscillates over an approximately constant range over the course considered here, except for the reaction of $p53$ degradation: the contribution of the $p53$ degradation accumulates over time. This observation is consistent with our propositions which predict for the steady-state models that the contribution of output signal degradation is significant, and generally overrides the contributions of other reactions. 
\par
In \cite{ito2010formulas} the authors studied the applicability of the LNA for the analysis of oscillatory models and found that the approximation fails in this type of models when long time periods are considered.  In these cases the total variance diverges to infinity with time. They also decomposed  the total variance into an oscillatory and a diverging part; the diverging part corresponds to the variability in the oscillation's period. The diverging part in Figure \ref{P53_contrib}  results from the p53 degradation, indicating that stochasticity of this reaction is responsible for the variability of the period in this system.

\section{Reducing the contribution of degradation noise}
\begin{figure}
\includegraphics[scale=0.4]{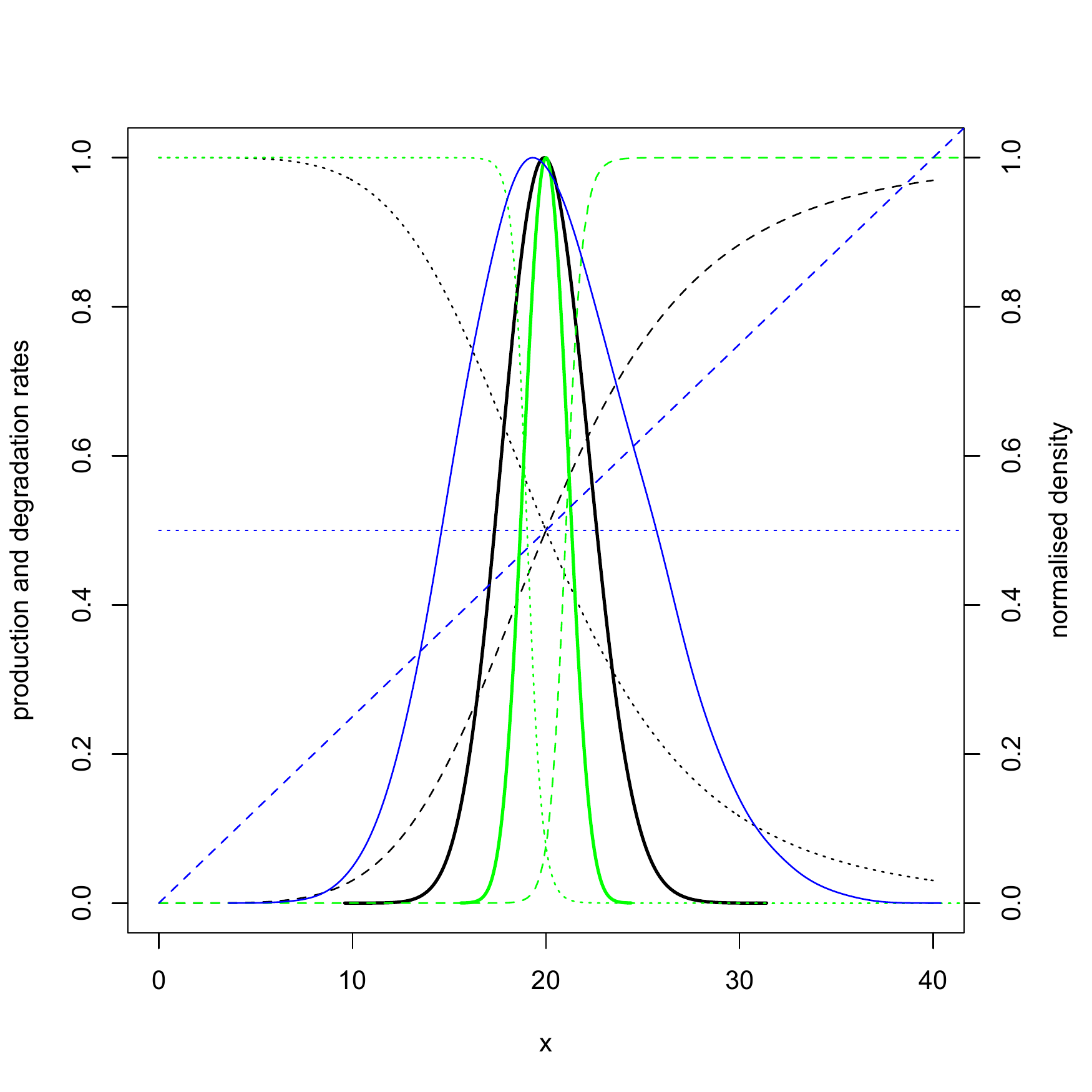}%
 \caption{ \label{Figure1} The effect of noise reduction resulting from regulated degradation. Each colour describes production rate $f(x)$ (dotted line) , degradation rate (dashed line) and density of stationary distribution of $x$ (solid line). For blue: $f(x)=1/2$, $g(x)=0.025x$; for  black: $f(x)=1/(1+(x/20)^5)$, $g(x)=(x/20)^5/(1+(x/20)^5)$; for green: $f(x)=1/(1+(x/19)^{50})$, $g(x)=(x/21)^{50}/(1+(x/22)^{50})$. Plotted densities are kernel density estimates based on 10000 independent stationary samples generated using Gillespie's algorithm \cite{Gillespie1977}. }
 \end{figure}

The propositions exemplified by the above models  demonstrate that the noise resulting from degradation of an output is a substantial source of variability in biochemical systems.  In a  general system it contributes a fraction of the variance equal to the half of the mean output. Here we show how controlling the degradation can reduce this contribution. We consider a simple birth and death process, but the same mechanism is valid for general system considered in the {\it Proposition 1}. If births and deaths occur at the state dependent rates $f(x)$ and $g(x)$, respectively, such a system is described by
\begin{equation} 
dx=(f(x)-g(x))dt+ \underbrace{\sqrt{f(\langle x \rangle)}dW_1}_{\text{birth noise}} +\underbrace{\sqrt{g(\langle x \rangle)}dW_2}_{\text{death noise}}.
\end{equation}
At equilibrium, we have $f(\langle x \rangle)=g(\langle x \rangle)$, and using formula (\ref{Sigma_j}) we can calculate that contribution of degradation is 
\begin{equation}
\Sigma^{(death)}= \frac{\frac{1}{2}g(\langle x \rangle)}{-\frac{\partial f(\langle x \rangle)}{\partial x}+\frac{\partial g(\langle x \rangle)}{\partial x}}.
\end{equation}
For $f(x)=k$, $g(x)= \gamma x$ this reduces, of course, to the previously discussed case $\Sigma^{(\text{death})}=\frac{1}{2}\langle x\rangle$. Nevertheless, if functions $f$ and $g$ are of Hill  and Michaelis-Menten type, respectively, then the contribution is no longer directly related to the mean and can be reduced to an arbitrary low level according to the above formula. %At equilibrium $f(\langle x \rangle)=g(\langle x \rangle)$ and the contributions of production and degradation to the variance are still equal to the half of the overall variance. 
Any decrease in the contribution compared to the mean is achieved by the reduction of the variance resulting from  the reduced flux through both reactions and autoregulatory control effects. Recent experimental work on protein degradation {\em in vitro} provides the evidence that degradation indeed exhibits Michaelis-Menten type kinetics \cite{gur2009degrons}, instead of the linear first order kinetics that are usually used to model degradation.  The effect of autoregulation is depicted in Figure \ref{Figure1}.

\section{Discussion}
The noise decomposition method introduced here allows us to investigate in detail where and how noise enters biochemical processes, is propagated through reaction systems, and affects cellular decision making processes.  We have shown analytically that in a wide class of systems the signal degradation reaction  contributes half of the noise in a system's output regardless of parameter values. We have also carried out numerical study for a system that never reaches steady-state and confirmed the surprisingly important role of degradation in controlling stochasticity of the p53 system, as well as in a range of generic systems biology models. Quite generally, the ability to dissect noise propagation through biological systems does enable researchers better to understand the role of noise in function (and evolution), and will also enable synthetic biologists to either harness or dampen the effects of noise in molecular signalling and response networks.
\par
One of the central results that we found and report here is the crucial role of degradation of the signal on the overall noise levels. 
The relevance of degradation has not been studied before in the context of stochastic biochemical dynamics and to our knowledge our study is the first that draws attention to the importance of degradation. This is particularly  important as this may indicate new therapeutic targets: in humans and other other sequenced organisms, certainly, the repertoire of proteins involved in protein degradation, in particular ubiquitin-ligases, is as rich and diverse as the repertoire of proteins regulating their activation, the kinases (and phosphatases). Thus targeting the degradation of proteins appears as important to biological systems as protein activation and offers an attractive and broad range of new potential therapeutic targets.

% link between deg and period of output signal
% maturation of FP 

\section{acknowledgments}
MK and MPHS acknowledge support from the BBSRC (BB/G020434/1).  
JM would like to thank the Polish Ministry of Science and Higher Education
for a financial support under the grant N201 362536.
MPHS is a Royal Society Wolfson Research Merit Award holder.

\appendix
\section{Proof of Proposition 1}
{
\noindent {\bf 1)} Interactions of $x_n$ with other species imply that $a_{ij}=0$ for $i=1,...,n-1$ and $j=n$. Thus matrix $A$ can be written as 
$$
A=\left(
\begin{array}{cc}
  \bar{A} &  0   \\
  a^T  &   -k^{-}_N \\
\end{array}
\right)
$$
\noindent {\bf 2)} Formula (\ref{D_j}) implies that all elements of matrix $D^{(r)}$ are equal to $0$ except $D^{(r)}_{nn}=k^{(-)}_n \langle x_n \rangle$, therefore $D^{(r)}$ has the form\
$$
D^{(r)}=\left(
\begin{array}{cc}
  0 &  0   \\
  0  &   k^{-}_n \langle x_n \rangle\\
\end{array}
\right)
$$
\noindent {\bf 3)} It is straightforward to verify that the $n\times n$ matrix 
\begin{equation}\label{SigR}
\Sigma^{(r)}=\left(
\begin{array}{cc}
  0 &  0   \\
  0  &   \frac{1}{2} \langle x_n \rangle  \\
\end{array}
\right)
\end{equation}
satisfies the equation 
\begin{equation}
{{A}}\Sigma^{(r)}+ \Sigma^{(r)} {{A}}^T + {{D}}^{(r)}=0
\end{equation}
}
\section{Proof of Proposition 2}
\noindent In an open conversion system $\Sigma_{nn}=\langle x_n \rangle$ (see e.g. \cite{gadgil2005stochastic}), therefore from {\it Proposition 1} we have $\Sigma_{nn}=2\Sigma^{(r)}_{nn}.$
% If you have acknowledgments, this puts in the proper section head.
%\clearpage

\bibliographystyle{unstr}

%\clearpage
  
\end{document}